\documentclass[12pt]{article}
\usepackage{graphicx}
\usepackage{psfrag}

\begin{document}

\begin{titlepage}
\null\vspace{-62pt}

\pagestyle{empty}
\begin{center}

\vspace{1.0truein} {\Large\bf Symmetry breaking and restoration for interacting scalar and gauge fields in Lifshitz type theories}

\vspace{1in}
{\large K.~Farakos and D.~Metaxas} \\
\vskip .4in
{\it Department of Physics,\\
National Technical University of Athens,\\
Zografou Campus, 15780 Athens, Greece\\
kfarakos@central.ntua.gr, metaxas@central.ntua.gr}\\

\vspace{0.5in}

\vspace{.5in}
\centerline{\bf Abstract}

\baselineskip 18pt
\end{center}

\noindent
We consider the one-loop effective potential at zero and finite temperature in field theories with anisotropic space-time scaling, with critical exponent $z=2$, including both scalar and gauge fields. Depending on the relative strength of the coupling constants for the gauge and scalar interactions, we find that there is a symmetry breaking term induced at one-loop at zero temperature and we find symmetry restoration through a first-order phase transition at high temperature.

\end{titlepage}

\newpage
\pagestyle{plain}
\setcounter{page}{1}
\newpage

Non-relativistic field theories in the Lifshitz context, with anisotropic scaling between temporal and spatial directions, measured by the dynamical critical exponent, $z$,
\begin{equation}
t\rightarrow b^z t,\,\,\,x_i\rightarrow b x_i,
\end{equation}
have been considered recently since they have an improved ultraviolet behavior and their renormalizability properties are quite different than conventional Lorentz symmetric theories \cite{visser}--\cite{jarev}.
Various field theoretical models and extensions of gauge field theories at the Lifshitz point have already been considered
\cite{hor3}.

When extended in curved space-time, these considerations may provide a renormalizable candidate theory of gravity \cite{hor1} and applications of these concepts in the gravitational and cosmological context
have also been widely investigated
\cite{kk}.

In order to investigate the various implications of a field theory, in particle physics and cosmology, it is particularly important to examine its symmetry structure, both at the classical and the quantum level, at zero and finite temperature, via the effective action and effective potential \cite{col}--\cite{farias}.
We should note that, in order to get information on possible instabilities of the theory, we study the one-loop, perturbative effective potential, given by the one-particle irreducible diagrams of the theory, and not the full, non-perturbative, convex effective potential given by the so-called Maxwell construction \cite{wett2}.

In a previous work \cite{before} we analyzed the case of a single scalar field and we found a symmetry breaking term induced at one-loop at zero temperature, as well as terms induced at finite temperature at one-loop, that provide symmetry restoration through a first-order phase transition. Because of the importance of symmetry breaking and restoration phenomena in quantum field theory and cosmology we extend here our previous work, including both scalar and gauge fields at the value $z=2$ of the critical exponent, and considering the physical implications of the several terms induced at one-loop in the effective potential at zero and finite temperature. We find that the results depend on the relative strength of the coupling constants for the scalar and the gauge fields, for a particular range of values of which we find similar effects as in \cite{before}; namely, when the strength of the gauge interaction is smaller than that of the scalar, in a way to be made clear from our result, we find symmetry breaking terms induced at one-loop at zero temperature, and symmetry restoration via terms that indicate a first-order phase transition at finite temperature.

We start with the action
\begin{eqnarray}
S=\int dt d^3x  \biggl(
-\frac{1}{2} F_{0i}F^{0i} +\frac{1}{4}F_{ij}\Delta F^{ij}
+\frac{1}{2}\left[ D_0\Phi\right]^{\dag}\left[ D_0\Phi\right] \nonumber\\
-\frac{1}{2}\left[D_i^2\Phi\right]^{\dag}\left[D_i^2\Phi\right] -U(\Phi) \biggr)
\end{eqnarray}
with $z=2$ where, as usual, $F_{\mu\nu}=\partial_{\mu}A_{\nu}-\partial_{\nu}A_{\mu}$,
$\Delta=\partial_i \partial_i$, $\Phi=\Phi_1+i\Phi_2$
and $D_{\mu}=\partial_{\mu}+ieA_{\mu}$.

The dimensionalities of the gauge fields and coupling are: $[A_0]=3/2$, $[A_i]=1/2$ and $[e]=1/2$.
The scalar field has
$[\Phi]=1/2$,
the potential term for $\Phi$ can, therefore, be an arbitrary polynomial up to the tenth order.
In order to illustrate the general features of the relative contributions of the gauge and scalar terms to the effective potential we will consider a potential term of the sixth order,
\begin{equation}
U(\Phi)=\frac{g}{6!}\left[\Phi^{\dag}\Phi\right]^3
\label{ss1}
\end{equation}
with $[g]=2$, where the strength of relative contributions, as it will turn out, will depend on the
dimensionless ratio $g/e^4$,
and we will also comment on the effects of other possible potential terms.

The action as it stands is gauge invariant under the $U(1)$ gauge transformations $A_{\mu}\rightarrow A_{\mu}+\partial_{\mu}\theta$, $\Phi\rightarrow e^{i\theta} \Phi$, and one can add a gauge-fixing term
\begin{equation}
L_{gf}=\frac{1}{2}(\partial_0 A_0 +\Delta\partial_i A_i)\frac{1}{\Delta}
(\partial_0 A_0 +\Delta\partial_i A_i)
\label{gft}
\end{equation}
via the usual Fadeev-Popov procedure, appropriately modified, in order to take care of the higher derivative and non-local terms.
Namely, one can impose the condition $G_f(A)=\partial_0 A_0 +\Delta\partial_i A_i=w(x)$,  which is an appropriate gauge-fixing condition for an arbitrary function $w(x)$, by inserting in the path integral an additional integration
\begin{equation}
\int[dw]\delta(G_f-w)e^{-\frac{i}{2}\int w\frac{1}{\Delta}w}.
\end{equation}
This has the effect of adding the previous term, (\ref{gft}), to the original, gauge-invariant action.

This gauge condition can also be derived via the BRST formalism \cite{ans}, and has the advantage of canceling the
mixed $A_0-A_i$ terms and making the calculation of the effective action more tractable.

We proceed by shifting the fields $\Phi_1\rightarrow\phi+\phi_1$,
$\Phi_2\rightarrow 0+\phi_2$, and now one can calculate the effective potential for $\phi$ at one-loop by the usual procedure \cite{jackiw}:
we write
\begin{equation}
L+L_{gf}=L_2 + L_{int},
\end{equation}
keeping quadratic order and $\phi$-dependent terms, where
\begin{equation}
L_2= \frac{1}{2}A_0 G^{-1} A_0 + \frac{1}{2}A_i \tilde{G}^{-1} A_i +
\frac{1}{2}\phi_1 K_1 \phi_1 +\frac{1}{2} \phi_2 K_2 \phi_2
\end{equation}
and
\begin{equation}
L_{int}=\frac{1}{2}e^2\phi^2 A_0^2 + e\phi A_0 \partial_0\phi_2
-\frac{1}{2}e^2\phi^2 (\partial_i A_i)^2 - e\phi(\Delta\phi_2)(\partial_i A_i).
\end{equation}
In momentum space the above terms are:
\begin{equation}
G^{-1}=\frac{k_0^2-k^4}{-k^2},
\end{equation}
\begin{equation}
\tilde{G}^{-1}=k_0^2-k^4,
\end{equation}
\begin{equation}
K_1=k_0^2-k^4-m_1^2(\phi),
\end{equation}
\begin{equation}
K_2=k_0^2-k^4-m_2^2(\phi).
\end{equation}
We use the notation: $k^2 =k_i^2$ for the spatial momentum and $m_1^2(\phi)=U''(\phi)$, $\phi m_2^2(\phi) = U'(\phi)$ for the
$\phi$-dependent masses. For the special potential term that we will analyze we have
$m_1^2(\phi) =\frac{1}{4!}g \phi^4$ and $m_2^2(\phi)= \frac{1}{5!} g \phi^4$, we will keep the general expression though, in order to give the most general result for the effective potential. It turns out that it is easier in terms of algebraic manipulations to perform the functional integration with respect to $A_0$ and $A_i$ first, then with respect to $\phi_2$, and finally with respect to $\phi_1$ which is straightforward.
The standard techniques of functional integration \cite{jackiw} give the final result for the effective potential at one-loop:
\begin{eqnarray}
U_{\rm eff}=&-&\frac{i}{2}Tr\ln K_1 -\frac{i}{2}Tr\ln K_2 -\frac{i}{2}Tr\ln D -\nonumber\\
&-&\frac{i}{2}Tr\ln\left(1+\frac{m_e^2(\phi)m_2^2(\phi)k^2}{(k_0^2-k^4)(k_0^2-k^4-m_2^2(\phi))}\right),
\label{effpot}
\end{eqnarray}
where $Tr=\int\frac{d^4 k}{(2\pi)^4}$, $m_e^2(\phi)=e^2\phi^2$ and
\begin{equation}
D=k_0^2-k^4-m_e^2(\phi)k^2
\end{equation}

The first two contributions in (\ref{effpot}), can be treated as in \cite{before}: using
\begin{equation}
\frac{d}{d(\alpha^2)}\int dk_0 \ln (k_0^2 + \alpha^2)=\frac{\pi}{\sqrt{\alpha^2}},
\end{equation}
integrating over $\alpha^2$
and dropping an overall constant, we get 
\begin{equation}
\int dk_0 \ln (k_0^2 + \alpha^2)=2\pi\sqrt{\alpha^2},
\end{equation}
so the first term in (\ref{effpot}) becomes
\begin{eqnarray}
-\frac{i}{2}Tr\ln K_1 &=&\frac{1}{4\pi^2}\int k^2 dk (k^4+m_1^2(\phi))^{1/2}=\nonumber\\
&=&\frac{1}{8\pi^2}m_1^2(\phi)\Lambda -c_1(m_1^2(\phi))^{5/4},
\end{eqnarray}
with
\begin{equation}
c_1=\frac{1}{4\pi^2}\int dx \frac{x^2}{(x^4+1)^{3/2}} =\frac{\Gamma(3/4)^2}{10\pi^{5/2}}
\end{equation}
after integrating with a momentum cutoff $\Lambda$.

The third term in (\ref{effpot}), that comes from the gauge field, can also be easily integrated with
a cutoff $\Lambda$, and we get
\begin{eqnarray}
-\frac{i}{2} Tr \ln D &=&\frac{1}{4\pi^2}\int k^2 dk (k^4 +m_e^2(\phi)k^2)^{1/2}=\nonumber\\
&=&\frac{1}{24\pi^2}m_e^2(\phi)\Lambda^3-\frac{1}{32\pi^2}m_e^4(\phi)\Lambda
+\frac{1}{30\pi^2}m_e^5(\phi).
\end{eqnarray}

We note that, as a consequence of the gauge-fixing condition chosen for this calculation, the divergence of this term that comes from the gauge field is cubic, higher than the divergence coming from the scalar fields. In this work we used a momentum cutoff, $\Lambda$; however, similar results can be derived using dimensional regularization \cite{farias}.

As far as the last, mixed contribution in (\ref{effpot}) is concerned, we can expand the logarithm and consider the first term that turns out to be finite:
\begin{eqnarray}
-\frac{i}{2}Tr \left(\frac{m_e^2(\phi)m_2^2(\phi)k^2}{(k_0^2-k^4)(k_0^2-k^4-m_2^2(\phi))}\right) =\nonumber\\
=\frac{m_e^2 m_2^2}{8\pi^3}\int dk_0dk\frac{k^4}{(k_0^2+k^4)(k_0^2+k^4+m_2^2)}=\nonumber\\
=\frac{m_e^2}{8\pi^3}\int dk_0 dk\,k^4\left(\frac{1}{k_0^2+k^4}-\frac{1}{k_0^2+k^4+m_2^2}\right)=\nonumber\\
=\frac{c_2}{4\pi^3}m_e^2(\phi)m_2^{3/2}(\phi),
\label{mix1}
\end{eqnarray}
where $c_2=\int dx dy \frac{x^4}{(y^2+x^4)(y^2+x^4+1)} = 0.970791$. It is easy to see that the higher order contributions coming from the expansion of the logarithm give terms that are also finite, all proportional to $\phi^5$, and include powers of the dimensionless ratio $e^4/g$. The symmetry breaking effects that appear in the effective potential come from the scalar loops, as we discuss below, and they dominate when this dimensionless ratio is small.
So, as a first approximation in terms of a coupling constant expansion, when $e^4/g << 1$ we will keep the first term that was calculated and write the complete expression for the effective potential at one-loop below. For the scalar potential term (\ref{ss1}) that we consider, we have:
\begin{eqnarray}
U_1(\phi)=\frac{g}{6!}\phi^6 &-&c_1\left(\frac{g}{4!}\phi^4\right)^{5/4}
-c_1\left(\frac{g}{5!}\phi^4\right)^{5/4}\nonumber\\
&+&\frac{(e\phi)^5}{30\pi^2}+\frac{c_2}{4\pi^3}e^2\phi^2\left(\frac{g}{5!}\phi^4\right)^{3/4}.
\label{effpot2}
\end{eqnarray}
A few comments are in order: first of all, the renormalization of the expression for the effective potential has been done by the addition of appropriate polynomial counterterms and imposing the conditions that the sixth derivative of the potential at $\phi=0$  is equal to $g$, and all the other derivatives are zero. For the particular choice of potential term that we consider, there is no infrared divergence at the origin as in \cite{before} and the renormalization is considerably simpler. The consideration of other potential terms is straightforward, the only difference being that some renormalization conditions for a general potential term have to be taken at a non-zero value of $\phi$.

We can also clearly see in this expression  that the relative strength of the contributions from the scalar and gauge fields depends on the dimensionless ratio $e^4/g$. The contribution of the scalar terms comes with a negative sign and has the effect of inducing symmetry breaking at one-loop as in \cite{before}. The contribution of the gauge and mixed terms, however, comes with a positive sign, and the final conclusion on the presence of symmetry breaking at one-loop depends on the dimensionless ratio of the scalar and gauge couplings.

An important point, that was first noticed and discussed in \cite{jackiw} is that of the gauge dependence of the results. It is well known that the effective potential, in fact the full effective action, in a theory that contains gauge fields depends on the gauge-fixing condition used. It is only the physical quantities that are inferred from this effective action that are gauge independent, as they should be. In a different gauge, the actual shape of the effective potential will be different; physical effects, however, such as the phenomenon of symmetry breaking, the value of the potential at the minimum, the tunneling rate from the false to the true vacuum are gauge independent. In \cite{jackiw} it was shown that one must use a renormalizable instead of a unitary gauge condition, and in \cite{nielsen} a general set of identities satisfied by the effective action were derived that demonstrate the gauge independence of the physical results in various applications \cite{metaxas}.

\vspace{0cm}

\begin{figure}[p]
\begin{center}

\includegraphics[scale=1]{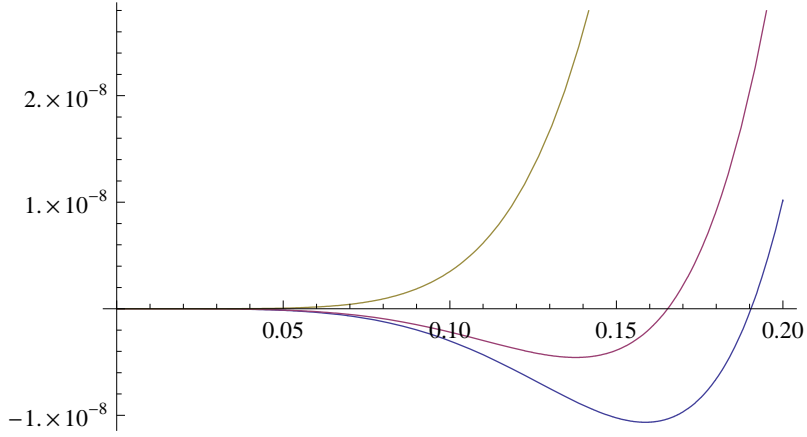}
\caption{ The full one-loop effective potential at zero temperature as a function of $\phi$
 for the scalar-gauge theory.
The potential is plotted in units of $\mu^5$ and $\phi$ is in units of
$\mu^{1/2}$.
With fixed $\tilde{g}=0.1$, the three curves shown correspond, from bottom to top, to $\tilde{e}=0.01, 0.02, 0.03$.}

\end{center}
\end{figure}

We plot the full expression for the effective potential in Fig.~1: the results are shown in terms of an overall dimensionful, ultraviolet scale, $\mu$. The potential is in units of $\mu^5$ and $\phi$ is in units of $\mu^{1/2}$. The coupling constants are $g=\tilde{g}\mu^2$ and $e=\tilde{e}\mu^{1/2}$. With fixed $\tilde{g}=0.1$ the three curves shown correspond to $\tilde{e}=0.01, 0.02, 0.03$. We see again the phenomenon of symmetry breaking induced by one-loop effects, which, however, disappears for increasing values of the gauge coupling constant, $e$.

\vspace{0cm}
\begin{figure}[p]
\begin{center}

\includegraphics[scale=1]{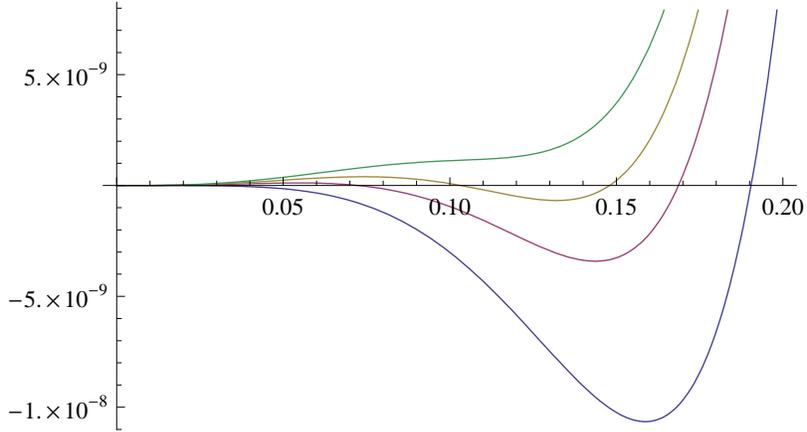}
\caption{ The full one-loop effective potential at finite temperature as a function of $\phi$
 for the scalar-gauge theory.
The potential is plotted in units of $\mu^5$, $\phi$ is in units of
$\mu^{1/2}$ and $T$ in units of $\mu^2$.
With fixed $\tilde{g}=0.1$ and  $\tilde{e}=0.01$ , the four curves shown correspond, from bottom to top, to temperatures $\tilde{T}=0, 0.002, 0.003, 0.004$.}

\end{center}
\end{figure}

The finite-temperature effective potential can now be calculated via the usual procedure of replacing
$\int\frac{d^4 k}{(2\pi^4)}\rightarrow T\sum_n \int\frac{d^3 k}{(2\pi^3)}$ and
$k_0\rightarrow i\omega_n$, where $\omega_n =2\pi n T$.
For the first three contributions in (\ref{effpot}) we use the identity
\begin{equation}
\sum_n\ln\left(\frac{4\pi^2n^2}{\beta^2}+E^2\right)=
2\beta\left[\frac{E}{2}+\frac{1}{\beta}\ln(1-e^{-\beta E})\right],
\end{equation}
in order to express them as a sum of the previous, one-loop, zero-temperature contribution and a temperature-dependent term. The last, mixed term of (\ref{effpot}) as it is written in (\ref{mix1}), is given at finite temperature by
\begin{equation}
\frac{m_e^2}{4\pi^2}T\int dk\,k^4 \sum_n\left(\frac{1}{\omega_n^2+k^4}-\frac{1}{\omega_n^2+k^4+m_2^2}\right)
\end{equation}
and can be split into the zero-temperature and temperature-dependent parts by using the identity
\begin{equation}
T\sum_n \frac{1}{\omega_n^2+E^2}=\frac{1}{2E}+\frac{1}{E(e^{\beta E}-1)}.
\end{equation}
Then the first, temperature-independent contributions combine to give the zero-temperature term of (\ref{mix1}) and the remaining terms give the temperature-dependent contribution of the mixed term.

Finally, the full potential can be written as the sum of the previous, one-loop, zero-temperature contribution, and a temperature-dependent part,
\begin{equation}
U=U_1+U_T,
\end{equation}
where $U_1$ is given by the previous expression (\ref{effpot2}) and
\begin{eqnarray}
U_T&=&T \int\frac{d^3 k}{(2\pi)^3} \ln \left( 1- e^{-\beta\sqrt{k^4+m_1^2(\phi)}} \right) \nonumber\\
   &+&T \int\frac{d^3 k}{(2\pi)^3} \ln \left( 1- e^{-\beta\sqrt{k^4+m_2^2(\phi)}} \right) \nonumber\\
   &+&T \int\frac{d^3 k}{(2\pi)^3} \ln \left( 1- e^{-\beta\sqrt{k^4+m_e^2(\phi)k^2}} \right)\nonumber\\
   &+&\frac{m_e^2}{4\pi^2}\int dk\left( \frac{k^2}{e^{\beta k^2}-1}-\frac{k^4}{\sqrt{k^4+m_2^2}(e^{\beta\sqrt{k^4+m_2^2}}-1)}\right)
\label{effpot3}
\end{eqnarray}
where, as usual $\beta=1/T$.
The first two terms in (\ref{effpot3}) are the contributions from the scalar fields and can be treated as in \cite{before}, with an analytical expression for the high temperature expansion:
\begin{equation}
T \int\frac{d^3 k}{(2\pi)^3} \ln \left( 1- e^{-\beta\sqrt{k^4+m_1^2(\phi)}} \right)
=
-\frac{\zeta(5/2)}{8 \pi^{3/2}}\, T^{5/2} +
\frac{2^{3/4}}{12\pi} T (m_1^2(\phi))^{3/4}.
\end{equation}
The third term in (\ref{effpot3}) is the contribution from the gauge fields and can be exactly evaluated as
\begin{equation}
T \int\frac{d^3 k}{(2\pi)^3} \ln \left( 1- e^{-\beta\sqrt{k^4+m_e^2(\phi)k^2}} \right)
=
-\frac{\zeta(5/2)}{8 \pi^{3/2}}\, T^{5/2} +
\frac{c_3}{4\pi^2}m_e^2(\phi) T^{3/2},
\end{equation}
with
$c_3=\int dx\frac{x^2}{e^{x^2}-1}= \frac{\sqrt{\pi}}{4}\zeta(3/2)=1.15758$.

The last, mixed term in (\ref{effpot3}) that corresponds to the temperature-dependent contribution of (\ref{mix1}) can be written after a rescaling as
\begin{equation}
\frac{m_e^2}{4\pi^2} T^{3/2}\int dx\left( \frac{x^2}{e^{x^2}-1}-\frac{x^4}{\sqrt{x^4+a^2}(e^{\sqrt{x^4+a^2}}-1)}\right),
\label{xxx1}
\end{equation}
where $a=\beta m_2$,
and can be given an analytical approximation at the high-temperature regime in the following manner:
first, we write
\begin{equation}
\int dx \, x^4 \frac{1}{\sqrt{x^4+a^2}(e^{\sqrt{x^4+a^2}}-1)}=
\frac{1}{a}\frac{\partial}{\partial a} \int dx \, x^4 \ln \left(1-e^{-\sqrt{x^4+a^2}}\right),
\end{equation}
then we expand the logarithm
\begin{equation}
\int dx \, x^4 \ln \left(1-e^{-\sqrt{x^4+a^2}}\right)=
-\sum_n \frac{1}{n}\int dx \, x^4 e^{-n\sqrt{x^4+a^2}}
\label{pp1}
\end{equation}
and use the high-temperature approximation of \cite{before}, $(x^4 +a^2)^{1/2}\approx x^2 + \frac{a^2}{2 x^2}$ (the integrand in (\ref{pp1}) is maximum for values of  $x\approx 1$ for small values of $n$ so our approximation to consider the main contribution for $x^2>a$ is valid since we are interested in the high-temperature regime $a<<1$).
Then the sum can be done with the help of the elementary integral
$\int dx e^{-A x^2- \frac{B}{x^2}}=\frac{1}{2}\sqrt{\frac{\pi}{A}}e^{-2\sqrt{A B}} $, expressed in terms of the Polylog function,
$P_\nu(w)=\sum_n \frac{1}{n^\nu} w^n$, with $w=e^{-\sqrt{2} a}$, and expanded in the high-temperature regime, where $a<<1$. The first terms of the expansion give
\begin{equation}
\int dx \, x^4 \frac{1}{\sqrt{x^4+a^2}(e^{\sqrt{x^4+a^2}}-1)}=
\frac{\sqrt{\pi}}{4}\zeta(3/2)-\frac{2^{1/4}\pi}{10} a^{1/2}+\cdots ,
\label{xxx2}
\end{equation}
the first term in this expression cancels the first term in (\ref{xxx1}) and
the final result for the temperature-dependent  contribution of the mixed term is
\begin{equation}
\frac{2^{1/4}}{40\pi}m_e^2(\phi)\,m_2^{1/2}(\phi) T.
\end{equation}

The effect of all the temperature-dependent contributions of the previous terms is to induce symmetry restoration at high temperature, via an apparently first-order phase transition, at least for values of the couplings, $g$ and $e$ that induce symmetry breaking effects at zero temperature, according to (\ref{effpot2}). In Fig.~2 we plot the full effective potential at finite temperature for $\tilde{g}=0.1$, $\tilde{e}=0.01$ for increasing temperatures. These values of the coupling give symmetry breaking effects at zero temperature, as was shown before, and the effective potential in Fig.~2 is plotted
for temperatures $\tilde{T}=0, 0.002, 0.003, 0.004$ where $T=\tilde{T}\mu^2$.
There is also a $\phi$-independent term in the expressions for the finite-temperature contribution that corresponds to the black-body radiation term (that is proportional to $T^4$ in the usual, Lorentz-invariant case) and is not shown in the figure. In our case this term is given by $\frac{3\zeta(5/2)}{8\pi^{3/2}}T^{5/2}$.
It is clear from these results that one has the interesting phenomenon of symmetry restoration at high temperature, with a potential term that indicates a first-order phase transition, for the small values of the dimensionless ratio $e^4/g$ which induce symmetry breaking at one-loop at zero temperature.

Several possible extensions of these results are possible; one may, for example, consider more general theories including gravitational and fermionic degrees of freedom, different values of the critical exponent, $z$, other than two, and, of course, it is also interesting to examine the general features and the development of the phase transition for field theories of the Lifshitz type.

\vspace{0.5in}

\centerline{\large\bf  Note} \noindent
After this work was completed we became aware of e-print \cite{farias} where some results similar to the ones obtained here were presented.


\vspace{0.5in}

\begin{thebibliography}{99}
\bibitem{visser} M.~Visser, {\it Phys. Rev.} {\bf D80}, 025011 (2009).
\bibitem{chen1} B.~Chen and Q.~G.~Huang, {\it Phys. Lett.} {\bf B683}, 108 (2010).
\bibitem{ans} D.~Anselmi and M.~Halat, {\it Phys. Rev.} {\bf D76}, 125011 (2007);
 D.~Anselmi, {\it Annals Phys.} {\bf 324}, 874 (2009); D.~Anselmi and M.~Taiuti, {\it Phys. Rev.} {\bf D81}, 085042 (2010).
\bibitem{iengo} R.~Iengo, J.~G.~Russo and M.~Serone, {\it JHEP} {\bf 0911}, 020 (2009).
\bibitem{jarev} J.~Alexandre, {\it Int. J. Mod. Phys.} {\bf A26}, 4523 (2011).
\bibitem{hor3} P.~Horava, {\it Phys. Lett.} {\bf B694}, 172 (2010);
 R.~Dijkgraaf, D.~Orlando and S.~Reffert, {\it Nucl. Phys.}
              {\bf B824}, 365 (2010);
J.~Alexandre and A.~Vergou, {\it Phys. Rev.} {\bf D83},
                 125008 (2011);
J.~Alexandre and N.~E.~Mavromatos, {\it Phys. Rev.} {\bf D83}, 127703 (2011);
A.~Dhar, G.~Mandal and S.~R.~Wadia, {\it Phys. Rev.} {\bf D80}, 105018 (2009);
J.~Alexandre, K.~Farakos, P.~Pasipoularides and A.~Tsapalis, {\it Phys. Rev.} {\bf D81}, 045002 (2010);
J.~Alexandre, K.~Farakos and A.~Tsapalis, {\it Phys. Rev.} {\bf D81}, 105029 (2010);
J.~E.~Thompson and R.~R.~Volkas, {\it Phys. Rev.} {\bf D82}, 116007 (2010);
K.~Anagnostopoulos, K.~Farakos, P.~Pasipoularides and A.~Tsapalis, arXiv:1007.0355.
\bibitem{hor1} P.~Horava, {\it JHEP} {\bf 0903}, 020 (2009);
P.~Horava, {\it Phys. Rev.} {\bf D79}, 084008 (2009);
P.~Horava and C.~M.~Melby-Thompson, {\it Phys. Rev.} {\bf D82}, 064027 (2010).

\bibitem{kk} E.~Kiritsis and G.~Kofinas, {\it Nucl. Phys.} {\bf B821}, 467 (2009);
S.~Mukohyama, {\it JCAP}, {\bf 0906}, 001 (2009);
R.~Brandenberger, {\it Phys. Rev.} {\bf D80}, 043516 (2009);
R.~G.~Cai, L.~M.~Cao and N.~Ohta, {\it Phys. Rev.} {\bf D80}, 024003 (2009);
S.~Mukohyama, K.~Nakayama, F.~Takahashi and S.~Yokoyama, {\it Phys. Lett.} {\bf B679}, 6 (2009);
 A.~Kehagias and K.~Sfetsos, {\it Phys. Lett.} {\bf B678}, 123 (2009);
C.~Charmousis, G.~Niz, A.~Padilla and P.~M.~Saffin, {\it JHEP} {\bf 0908}, 070 (2009);
G.~Koutsoumbas and P.~Pasipoularides, {\it Phys. Rev.} {\bf D82}, 044046 (2010);
 M.~Eune and W.~Kim, {\it Mod. Phys. Lett.} {\bf A25}, 2923 (2010);
D.~Orlando and S.~Reffert, {\it Class. Quant. Grav.} {\bf 26}, 155021 (2009);
 M.~Jamil, E.~N.~Saridakis and M.~R.~Setare, {\it JCAP} {\bf 1011}, 032 (2010);
G.~Koutsoumbas, E.~Papantonopoulos, P.~Pasipoularides and M.~Tsoukalas, {\it Phys. Rev.} {\bf D81}, 124014 (2010);
 C.~Soo, J.~Yang and H.~L.~Yu, {\it Phys. Lett.} {\bf B701}, 275 (2011);
J.~Alexandre and P.~Pasipoularides, {\it Phys. Rev.} {\bf D83}, 084030 (2011).

\bibitem{col} S.~R.~Coleman and E.~J.~Weinberg, {\it Phys. Rev.} {\bf D7}, 1888 (1973).
\bibitem{jackiw} R.~Jackiw, {\it Phys. Rev.} {\bf D9}, 1686 (1974).
\bibitem{bran2} R.~H.~Brandenberger, {\it Rev. Mod. Phys.} {\bf 57}, 1 (1985).
\bibitem{dolan} L.~Dolan and R.~Jackiw, {\it Phys. Rev.} {\bf D9}, 3320 (1974).
\bibitem{before} D.~Metaxas and K.~Farakos, arXiv:1109.0421.
\bibitem{wett2} C.~Wetterich, {\it Nucl. Phys.} {\bf B352}, 529 (1991); J.~Alexandre, arXiv:0904.0934[hep-ph].
\bibitem{kim1} M.~Eune, W.~Kim and E.~J.~Son, {\it Phy. Lett.} {\bf B703}, 100 (2011).
\bibitem{lopez} D.~L.~Nacir, F.~D.~Mazzitelli and L.~G.~Trombetta, arXiv:1111.1662.
\bibitem{farias} C.~F.~Farias, M.~Gomez, J.~R.~Nascimento, A.~Y.~Petrov and A.~J.~da~Silva, arXiv:1112.2081.
\bibitem{nielsen} N.~K.~Nielsen, {\it Nucl. Phys.} {\bf B101}, 173 (1975).
\bibitem{metaxas} R.~Fukuda and T.~Kugo, {\it Phys. Rev.} {\bf D13}, 3469 (1976);
I.~J.~R.~Aitchison and C.~M.~Fraser, {\it Annals Phys.} {\bf 156}, 1 (1984);
D.~Metaxas and E.~J.~Weinberg, {\it Phys. Rev.} {\bf D53}, 836 (1996);
D.~Metaxas, {\it Phys. Rev.} {\bf D63}, 085009 (2001).

\end{thebibliography}
\end{document}